\documentclass[preprint,12pt]{elsarticle}

\usepackage{amssymb}
\usepackage{amsmath}
\usepackage{siunitx}
\usepackage{makecell}
\usepackage{natbib}
\usepackage{placeins}

\journal{Green Energy \& Environment}

\begin{document}

\begin{frontmatter}

%% Title, authors and addresses

%% use the tnoteref command within \title for footnotes;
%% use the tnotetext command for theassociated footnote;
%% use the fnref command within \author or \address for footnotes;
%% use the fntext command for theassociated footnote;
%% use the corref command within \author for corresponding author footnotes;
%% use the cortext command for theassociated footnote;
%% use the ead command for the email address,
%% and the form \ead[url] for the home page:
%% \title{Title\tnoteref{label1}}
%% \tnotetext[label1]{}
%% \author{Name\corref{cor1}\fnref{label2}}
%% \ead{email address}
%% \ead[url]{home page}
%% \fntext[label2]{}
%% \cortext[cor1]{}
%% \affiliation{organization={},
%%             addressline={},
%%             city={},
%%             postcode={},
%%             state={},
%%             country={}}
%% \fntext[label3]{}

%\title{Carbon Trapping Efficiency by Hydropower Reservoirs under Tropical Climate Influence}

\title{Carbon Trapping Efficiency of Hydropower Reservoirs under the Influence of a Tropical Climate}

%% use optional labels to link authors explicitly to addresses:
%% \author[label1,label2]{}
%% \affiliation[label1]{organization={},
%%             addressline={},
%%             city={},
%%             postcode={},
%%             state={},
%%             country={}}
%%
%% \affiliation[label2]{organization={},
%%             addressline={},
%%             city={},
%%             postcode={},
%%             state={},
%%             country={}}

%%%%%%%%%%%
% AUTHORS %
%%%%%%%%%%%

% \author[inst1]{Marco Aurélio dos Santos}
% \affiliation[inst1]{organization={Energy Planning Program/COPPE/UFRJ},%Department and Organization
%             addressline={Centro de Tecnologia, Bloco C, sala 211, Cidade Universitária}, 
%             city={Rio de Janeiro},
%             country={Brazil}}
% \author[inst2]{Bohdan Matvienko}
% \affiliation[inst2]{organization={CHREA/USP and Chemistry of Construmaq São Carlos}}
% \author[inst3]{Elisabeth Sikar}
% \affiliation[inst3]{organization={Physics of Construmaq São Carlos}}
% \author[inst4]{Jorge Machado Damázio}
% \affiliation[inst4]{organization={Center of Electric Research – CEPEL/Eletrobras}}
% \author[inst1]{Marcelo Andrade Amorim}
% \author[inst5]{Marcela Vidal}
% \affiliation[inst5]{organization={Geoquemistry Department/UFF}}
% \author[inst5]{Marcos Manoel Ferreira}
% \author[inst6]{Karen de Jesus}
% \affiliation[inst6]{organization={Instituto de Química de São Carlos, Universidade de São Paulo-USP}}
% \author[inst1]{Gustavo Couto}

%\author[1]{Marco Aurélio dos Santos}
\author[1]{Marco Aurélio dos Santos\corref{cor1}}
\ead{aurelio@ppe.ufrj.br}
\cortext[cor1]{Corresponding author}
\author[2]{Bohdan Matvienko}
\author[3]{Elizabeth Sikar}
\author[4]{Jorge Machado Damázio}
\author[1]{Marcelo Andrade Amorim}
\author[5]{Marcela Vidal}
\author[5]{Marcos Manoel Ferreira}
\author[6]{Karen de Jesus}
\author[1]{Gustavo Couto}
\author[7]{Daniel Sikar}

\affiliation[1]{organization={Energy Planning Program/COPPE/UFRJ, Centro de Tecnologia, Bloco C, sala 211, Cidade Universitária, Rio de Janeiro, Brazil}}
\affiliation[2]{organization={Professor at CRHEA/USP and Chemist at Construmaq São Carlos - "in memoriam"}}
\affiliation[3]{organization={Physicist at Construmaq São Carlos – emsikar@gmail.com}}
\affiliation[4]{organization={Researcher at the Center for Electric Research– CEPEL/Eletrobras – damazio@cepel.br}}
\affiliation[5]{organization={Doctoral Student - Geochemistry Department/UFF}}
\affiliation[6]{organization={Instituto de Química de São Carlos, Universidade de São Paulo-USP}}
\affiliation[7]{organization={Software Developer at Construmaq São Carlos – dsikar@gmail.com}}
\date{}
% \author[inst1]{Author One}

% \affiliation[inst1]{organization={Department One},%Department and Organization
%             addressline={Address One}, 
%             city={City One},
%             postcode={00000}, 
%             state={State One},
%             country={Country One}}

% \author[inst2]{Author Two}
% \author[inst1,inst2]{Author Three}

% \affiliation[inst2]{organization={Department Two},%Department and Organization
%             addressline={Address Two}, 
%             city={City Two},
%             postcode={22222}, 
%             state={State Two},
%             country={Country Two}}

\begin{abstract}
Sedimentation in hydroelectric reservoirs is strongly impacted by anthropogenic activities within their upstream drainage basins. These activities, encompassing soil erosion and various other human-induced actions, have significant consequences for sedimentation patterns.

This issue has been a subject of prolonged study, as sedimentation directly undermines the water storage capacity of reservoirs, consequently diminishing the overall efficiency of hydroelectric operations.

Several scientists have dedicated their efforts to addressing the matter of reservoir sedimentation. This pursuit has led to the formulation of an indicator known as Sediment Trap Efficiency (STE), serving as a metric that quantifies the proportion of sedimentation within reservoirs relative to the influx of sediment from their upstream sources.

This study seeks to present findings pertaining to carbon trapping efficiency observed across seven hydroelectric reservoirs in Brazil. The objective is to demonstrate the substantial relevance of carbon accumulation within these aquatic environments within the context of the carbon balance frameworks previously established.
\end{abstract}

%%Graphical abstract
\begin{graphicalabstract}
\includegraphics[width=0.99\columnwidth]{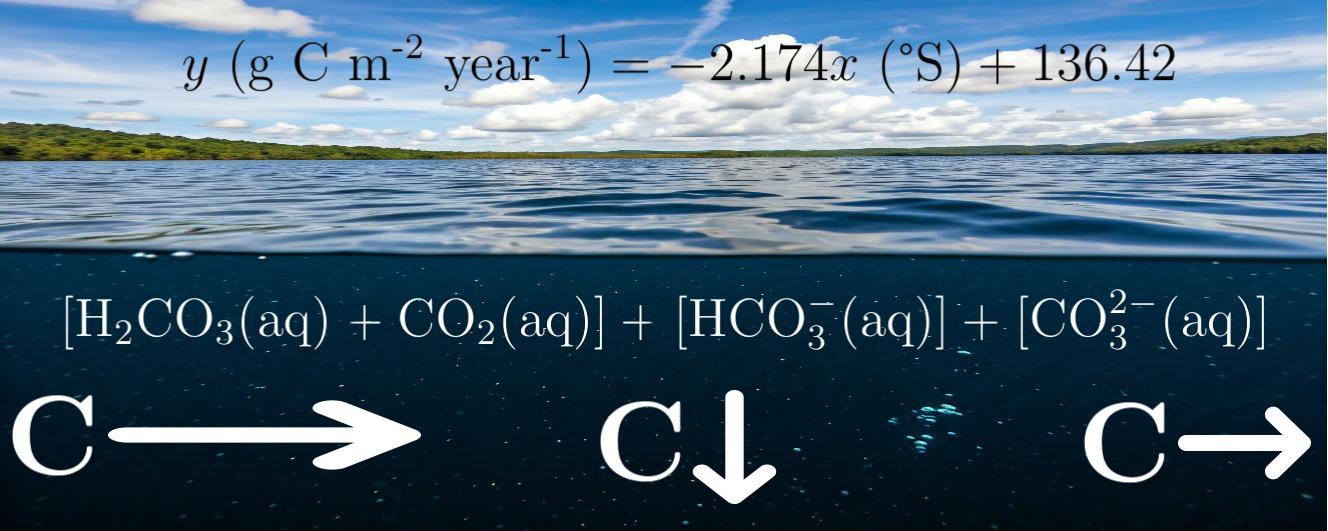}
\end{graphicalabstract}

%%Research highlights
\begin{highlights}
\item Carbon burial rates in the studied Brazilian hydropower reservoirs range from 67.38 to 196.14 g C m$^{-2}$ year$^{-1}$, demonstrating significant carbon sequestration potential

\item An inverse relationship exists between carbon burial rates and latitude (1.9°S to 25.8°S), described by the equation: y = -2.174x + 136.42 (R$^2$ = 0.2496)

\item Carbon trapping efficiency varies widely (3.66\% to 48.6\%) among reservoirs, with a trend of decreasing efficiency by 0.994\% per degree increase in latitude

\item Reservoirs in unaltered watersheds (rainforest and caatinga) show higher variability in carbon influx ratios (0.0003 to 0.08 t C km$^{-2}$ d$^{-1}$) compared to anthropogenically impacted watersheds (0.01 to 0.02 t C km$^{-2}$ d$^{-1}$)

\item The study demonstrates that tropical and subtropical hydropower reservoirs can serve as substantial carbon sinks, with implications for climate change mitigation strategies
\end{highlights}

\begin{keyword}
%% keywords here, in the form: keyword \sep keyword
%keyword one \sep keyword two
%\keywords{Carbon trapping, Hydropower reservoirs, Tropical climate, Sedimentation}
Carbon trapping \sep Hydropower reservoirs \sep Tropical climate \sep Sedimentation

%% PACS codes here, in the form: \PACS code \sep code
% https://publishing.aip.org/wp-content/uploads/2019/01/PACS_2010_Alpha.pdf
\PACS 89.60 \sep  88.05
% https://cran.r-project.org/web/classifications/MSC.html
%% MSC codes here, in the form: \MSC code \sep code
%% or \MSC[2008] code \sep code (2000 is the default)
%\MSC 0000 \sep 1111
\end{keyword}

\end{frontmatter}

%% \linenumbers

%% main text
% INTRODUCTION
%%%%%%%%%%%%%%%%
% INTRODUCTION %
%%%%%%%%%%%%%%%%

\section{Introduction}

Sedimentation in hydroelectric reservoirs is strongly impacted by anthropogenic activities within their upstream drainage basins. These activities, encompassing soil erosion and various other human-induced actions, have significant consequences for sedimentation patterns.

This issue has been under investigation for an extended duration, given that sedimentation diminishes the water storage capacity of reservoirs and consequently hampers the efficiency of hydroelectric utilization.

Citing \cite{friedl2002disrupting}, over 800,000 lakes and artificial reservoirs have been constructed globally for diverse purposes, such as human water supply, hydroelectric power generation, irrigation, and other applications.

The dams built and their respective reservoirs have increased the accumulation of sediments along the rivers, preventing the material from reaching the oceans. Recent publications have estimated that the retention of suspended particles in reservoirs can reach about 50\% of the load that would reach the oceans \cite{vorosmarty2003anthropogenic, walling2006human}.

Global estimates were performed to demonstrate that the reduction of water storage capacity may be a future problem. A report by the World Commission on Dams estimated that the loss of water accumulation from sedimentation of reservoirs varies at a rate of between 0.5\% to 1\% per year \cite{wcd2000dams}.

Studies carried out by \cite{wisser2013beyond}, based on computer simulations, showed that between the period from 1901 to 2010, the world average of water storage in reservoirs was reduced by 5\%.

Numerous researchers have investigated the matter of reservoir sedimentation, leading to the development of an indicator known as Sediment Trap Efficiency (STE). This index gauges the sedimentation rate within reservoirs relative to the influx of sediments from upstream sources \cite{vorosmarty2003anthropogenic, kummu2010basin, obreja2012assessment, yang2014estimate}.

The efficiency of sediment retention depends on some characteristics such as the type and amount of sediments that reach these reservoirs, as well as the hydraulic retention time of these water bodies, controlled by the geometry and flow characteristics of the drainage basin \cite{verstraeten2000estimating}.

On the other hand, a series of studies have also been carried out to propose measures to manage and reduce the problem \cite{kondolf2014sustainable}.

Running in tandem with studies on sedimentation, research into the interplay between deposited carbon and carbon sequestered within sediments has been stimulated. This exploration encompasses an examination of the efficacy of carbon sequestration in lakes and reservoirs, aimed at countering the natural reintegration of this carbon through its permanent entrenchment within the sedimentary layers of these aquatic systems.

In this context, the terms "permanent carbon sedimentation," "carbon burial," "carbon sequestration," and "carbon fixation" are used interchangeably. Furthermore, carbon trapping efficiency is defined as the ratio of carbon sequestered within the sediment to the incoming carbon inflow from tributaries into the reservoir.

The primary objective of this study is to offer a comprehensive understanding of carbon trapping efficiency in seven hydroelectric reservoirs in Brazil, emphasizing the significant role of carbon sequestration within these aquatic systems in the context of established carbon balances.
% RELATED WORK
%%%%%%%%%%%%%%%%%%%%
% PREVIOUS RESULTS %
%%%%%%%%%%%%%%%%%%%%

\section{Previous Results on Carbon Sedimentation}

Various studies have been conducted to assess the significance of freshwater reservoirs in sequestering carbon within sediment deposits on the lakebed. These reservoirs serve as crucial carbon sinks, as evidenced by research conducted by \cite{mulholland1982role}, \cite{dean1998magnitude}, \cite{martynova2006production}, \cite{cole2007plumbing}, \cite{sobek2009organic}, \cite{kunz2011sediment}, \cite{mendonca2012hydroelectric}, \cite{teodoru2012net}, \cite{knoll2013temperate}, and \cite{mendonca2014carbon}.

In a published work, \cite{li2014dam} assert that artificial reservoirs have played a significant role in the retention of carbon, particularly in regard to the particulate organic carbon fraction.

The study conducted by \cite{mulholland1982role} proposed a hypothesis suggesting that artificial reservoirs could potentially function as carbon sinks under certain conditions. Specifically, they posited that if the carbon sedimented at the reservoir bottoms constituted a portion of the carbon introduced via inflowing rivers, or if this carbon would otherwise undergo oxidation and atmospheric release in pre-damming conditions (prior to reservoir formation). The authors build upon a series of previous investigations conducted in 51 reservoirs worldwide, culminating in the determination of an average carbon accumulation value of $500 \, \text{g} \, \text{C} \, \text{m}^{-2} \, \text{year}^{-1}$ in artificial reservoirs.

In the study conducted by \cite{knoll2013temperate}, which assessed the carbon balance in two freshwater reservoirs under distinct land use conditions (forestry and agriculture), a carbon accumulation rate ranging from $274$ to $340 \, \text{g} \, \text{C} \, \text{m}^{-2} \, \text{year}^{-1}$ was observed (Acton Reservoir and Bur Oak Reservoir).

\cite{almeida2016high} suggested that, excluding the mineralization of organic forms of carbon from the sediment at the base of a small reservoir in the semi-arid region of the caatinga biome in Brazil, the sedimentation rate is approximately $440 \, \text{mg} \, \text{C} \, \text{m}^{-2} \, \text{day}^{-1}$. This corresponds to approximately $160.6 \, \text{g} \, \text{C} \, \text{m}^{-2} \, \text{year}^{-1}$, a value in line with the previously presented findings.

\cite{sobek2009organic} and collaborators studied 11 lakes around the world and concluded that the rate of organic carbon buried in bottom sediments ranged from $0.2$ to $1,140 \, \text{g} \, \text{C} \, \text{m}^{-2} \, \text{year}^{-1}$, with a median value of $27$ and an average of $192 \, \text{g} \, \text{C} \, \text{m}^{-2} \, \text{year}^{-1}$.

\cite{martynova2006production} and collaborators raised results of sedimented carbon studies in 11 freshwater reservoirs in the world, with an average rate of $59 \, \text{g} \, \text{C} \, \text{m}^{-2} \, \text{year}^{-1}$ and in 23 freshwater lakes a rate of $65.1 \, \text{g} \, \text{C} \, \text{m}^{-2} \, \text{year}^{-1}$.

According to \cite{dean1998magnitude}, freshwater lakes and artificial reservoirs deposit more organic carbon at the bottom of their bodies of water annually than the oceans as a whole. This study estimates that buried carbon in these freshwater bodies ranges from $0.16$ to $0.2 \, \text{Pg} \, \text{C} \, \text{year}^{-1}$ with an average value of $0.18 \, \text{Pg} \, \text{C} \, \text{year}^{-1}$. The study further notes that the $0.16 \, \text{Pg} \, \text{C} \, \text{year}^{-1}$ value corresponds to a rate of $400 \, \text{g} \, \text{C} \, \text{m}^{-2} \, \text{year}^{-1}$.

The initial years or decades of an artificial reservoir typically exhibit elevated rates of carbon sedimentation.

Upon utilizing the surface area measurements of large reservoirs as estimated by Saint Louis and colleagues \cite{st2000reservoir}, 
\cite{cole2007plumbing} determined that the estimate put forth by Dean and Gorham in 1998 is indeed conservative. The reevaluated figure could potentially reach $0.6 \, \text{Pg} \, \text{year}^{-1}$.

These recent estimations postulated that the rates of organic carbon accumulation in lakes and reservoirs surpass the rates of deposition in the ocean cite{cole2007plumbing, williamson2009lakes}.

Carbon accumulation rates derived from sedimentation traps were established by \cite{teodoru2012net}, yielding an average of $90 \, \text{mg} \, \text{C} \, \text{m}^{-2} \, \text{day}^{-1}$, with values ranging between $30$ and $142 \, \text{mg} \, \text{C} \, \text{m}^{-2} \, \text{day}^{-1}$. The study notes that carbon sedimentation rates in the Eastmain 1 hydroelectric reservoir in Canada are three to four times greater than those observed in nearby natural lakes.
% METHODS
%%%%%%%%%%%%%%%%%%%%
% SITE AND METHODS %
%%%%%%%%%%%%%%%%%%%%

%\section{Sites and Methods}
\section{Theory and Calculations}

\subsection{Site Selection}

Seven hydroelectric reservoirs were investigated across the expanse of Brazilian territory, including a diverse range of technical characteristics within the projects. These include factors such as the power capacity of the enterprise, reservoir area, the extent of the drainage basin upstream of the reservoir, and the specific biome type in which the projects are situated.

\begin{figure}[ht]
    \centering
    \includegraphics[width=0.99\columnwidth]{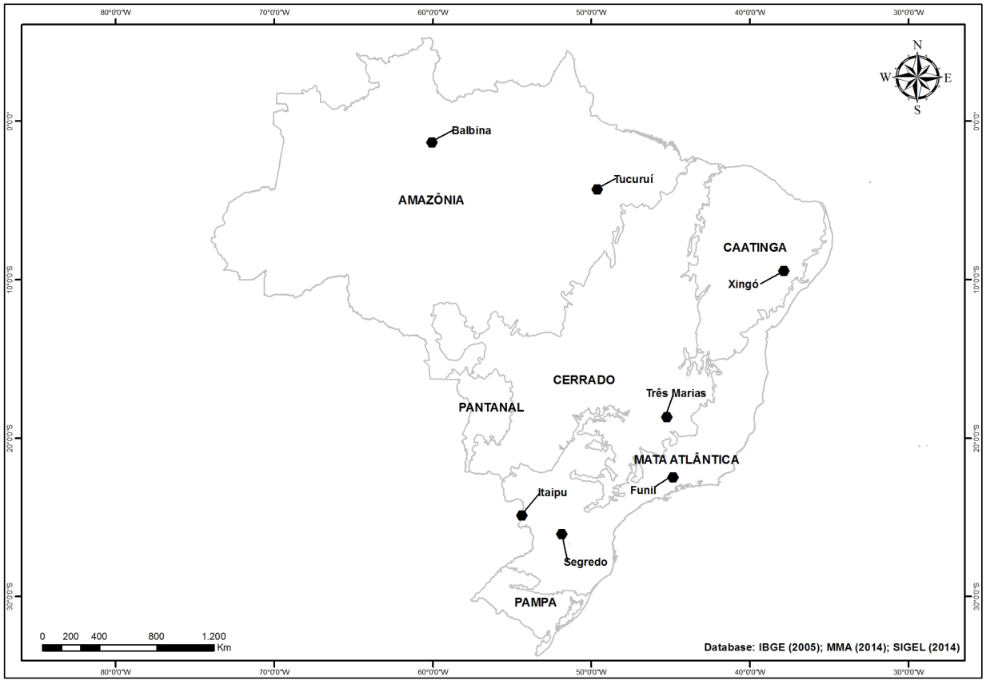}
    \caption{Geographic Location of the Studied Reservoirs}
    \label{fig:geo_location}
\end{figure}

Table \ref{tab:reservoir-characteristics} below presents the key information regarding the studied projects, while Figure \ref{fig:geo_location} shows the geographical locations of the enterprises under examination.

% \begin{table}
% \footnotesize
% \centering
% \caption{Characteristics of the Studied Reservoirs}
% \label{tab:reservoir-characteristics}
% \begin{tabular}{lllrrrr}
% \hline
% Reservoir & Latitude & Biome & Power (MW) & \makecell{Reservoir\\Area (km$^2$)} & \makecell{Energy\\Density (W/m$^2$)} & \makecell{Upstream\\Watershed\\Area (km$^2$)} \\
% \hline
% Três Marias & 18°13'S & Cerrado & 396 & 1,040 & 0.38 & 38,136.06 \\
% Tucuruí & 3°45'S & Amazon Forest & 8,000 & 3,000 & 2.66 & 759,237.48 \\
% Balbina & 1°53'S & Amazon Forest & 250 & 2,360 & 0.1 & 14,207 \\
% Segredo & 25°47'S & Atlantic Forest & 1,260 & 82 & 15.37 & 34,267.95 \\
% Xingó & 9°37'S & Caatinga & 3,000 & 60 & 50 & 614,608.74 \\
% Itaipu & 24°43'S & Atlantic Forest & 12,600 & 1,546 & 8.15 & 803,134.71 \\
% Funil & 22°31'S & Atlantic Forest & 216 & 40 & 5.4 & 12,781.91 \\
% \hline
% \end{tabular}
% \end{table}

%% SPLIT TABLES %%

\begin{table}[h!]
%\footnotesize
\centering
\caption{Characteristics of the Studied Reservoirs}\vspace{2pt}
\label{tab:reservoir-characteristics}
\begin{tabular}{lllrr}
\hline
Reservoir & Latitude & Biome & Power (MW) & \makecell{Reservoir \\ Area (km$^2$)} \\
\hline
Três Marias & 18°13'S & Cerrado & 396 & 1,040 \\
Tucuruí & 3°45'S & Amazon Forest & 8,000 & 3,000 \\
Balbina & 1°53'S & Amazon Forest & 250 & 2,360 \\
Segredo & 25°47'S & Atlantic Forest & 1,260 & 82 \\
Xingó & 9°37'S & Caatinga & 3,000 & 60 \\
Itaipu & 24°43'S & Atlantic Forest & 12,600 & 1,546 \\
Funil & 22°31'S & Atlantic Forest & 216 & 40 \\
\hline
\end{tabular}
\end{table}

\FloatBarrier % This prevents any floats from crossing this point

\begin{table}[h!]
%\footnotesize
\centering
\caption{Characteristics of the Studied Reservoirs - continued}\vspace{2pt}
\label{tab:reservoir-characteristics-cont}
\begin{tabular}{lrr}
\hline
Reservoir & \makecell{Energy Density (W/m$^2$)} & \makecell{Upstream Watershed Area (km$^2$)} \\
\hline
Três Marias & 0.38 & 38,136.06 \\
Tucuruí & 2.66 & 759,237.48 \\
Balbina & 0.1 & 14,207 \\
Segredo & 15.37 & 34,267.95 \\
Xingó & 50 & 614,608.74 \\
Itaipu & 8.15 & 803,134.71 \\
Funil & 5.4 & 12,781.91 \\
\hline
\end{tabular}
\end{table}

Four field campaigns were conducted in each reservoir between 2011 and 2013. These campaigns aimed to incorporate seasonal variations of the climatic conditions at the sampling sites into the studies.

Table \ref{tab:sampling-points} presents the count of sampling sites within each studied hydroelectric reservoir.

\begin{table}[h!]
\centering
\caption{Count of Sampled Sites in Each Reservoir}
\label{tab:sampling-points}
\begin{tabular}{lcccc}
\hline
Reservoir & 1st Campaign & 2nd & 3rd & 4th \\
\hline
Balbina & 12 & 17 & 15 & 16 \\
Funil & 11 & 10 & 11 & - \\
Itaipu & 11 & 21 & 21 & 12 \\
Segredo & 21 & 21 & 21 & 21 \\
Tres Marias & 17 & 20 & 19 & 16 \\
Tucuruí & 23 & 23 & 21 & 21 \\
Xingó & 17 & 15 & 13 & 13 \\
\hline
\end{tabular}
\end{table}

\subsection{Methods}

\subsubsection{Carbon Sedimentation}

Rates of permanent carbon sedimentation were established utilizing silicon as a tracer, as outlined by \cite{sikar2012silicon}. This approach necessitates three measurements to derive the permanent carbon sedimentation rate (C), quantified in mg C m\textsuperscript{-2} d\textsuperscript{-1}: the sedimentation rate of silicon (mg Si m\textsuperscript{-2} d\textsuperscript{-1}), the silicon concentration [Si] within the sediment profile (\% Si), and the concentration of carbon [C] within the sediment profile (\% C).

After the sedimentation rate (T) of Si is determined using sediment traps, and the ratio (R) between [C] and [Si] is established within the enduring sediment layer, the sedimentation rate for permanent carbon becomes:

\begin{equation}
P = T \times R
\label{eq:permanent_carbon}
\end{equation}

While not strictly mandatory, an additional measurement involves assessing the daily sedimentation rate of 'fresh carbon' (C\textsubscript{f}), referring to carbon that descends through the water column to a depth of approximately 1 meter above the sediment. This particular range constitutes a pivotal region where the carbon's fate—whether it undergoes decomposition or fossilization—is ultimately determined. This evaluation, facilitated through trap utilization, checks the Si-tracer methodology. The rationale for this approach stems from the necessity for the daily sedimentation rate of permanent carbon to ideally remain lower than the daily rate of C\textsubscript{f} sedimentation. This underscores the importance of maintaining coherence between these two methods.

The method for sample collection and analysis is outlined in a prior publication authored by some of the researchers of this study \cite{sikar2012silicon}.

Sedimentation traps are made of 40 cm long ($l$) polyvinyl chloride (PVC) tubes, 7.1 cm in diameter ($d$) and closed at the bottom. The aspect ratio ($l/d$) 5.6 should be used to minimize the interference of the trap in the measurements \cite{rosa1994sampling}.

\begin{figure}
    \centering
    \includegraphics[width=0.5\textwidth]{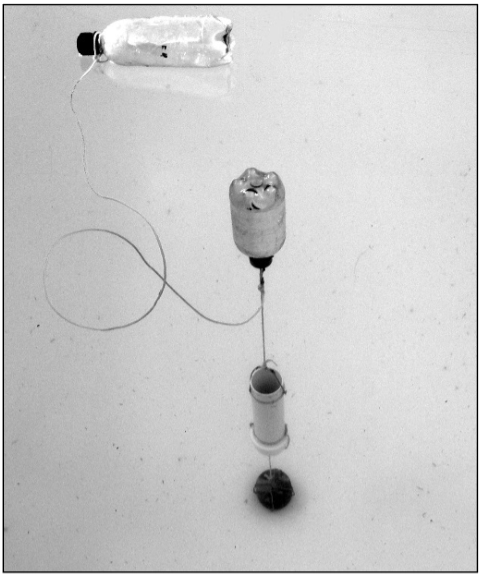}
    \caption{Sedimentation trap}
    \label{fig:sedimentation-trap}
\end{figure}

When placed in the water, a sling is connected to the bottom of the trap by a short ($\sim$0.5 m) rope. The top of the trap is attached to a suspension rope. A mockup image of this arrangement is shown in Figure \ref{fig:sedimentation-trap}.

\subsubsection{Quantifying Carbon Mass Inflow from Tributaries to Reservoir}

During each campaign, which served a range of objectives beyond carbon sedimentation sampling, the influx of allochthonous carbon into the reservoirs via the tributaries was determined. This involved conducting sampling to quantify the inflow of carbon mass from all tributaries connected to the studied reservoirs.

Flow measurement quantifies the volume of water that passes through a given section of a channel in a given period of time. This method depends on the knowledge of physical variables such as width, depth and flow velocity, according to the International System (SI) of measurements described in \cite{chevallier2009aquisicao}.

The flow is quantified in cubic meters per second (m\textsuperscript{3} s\textsuperscript{-1}) and expressed mathematically by the equation below:

\begin{equation}
Q = (w \cdot h) \cdot V
\label{eq:flow}
\end{equation}

Where:
\begin{itemize}
\item Q = flow (m\textsuperscript{3} s\textsuperscript{-1})
\item A = river section area (m\textsuperscript{2}) (w·h)
\item V = water flow velocity (m s\textsuperscript{-1})
\item h = average depth in the cross section of the channel (m)
\item w = channel width (m)
\end{itemize}

\begin{figure}
    \centering
    \includegraphics[width=0.99\textwidth]{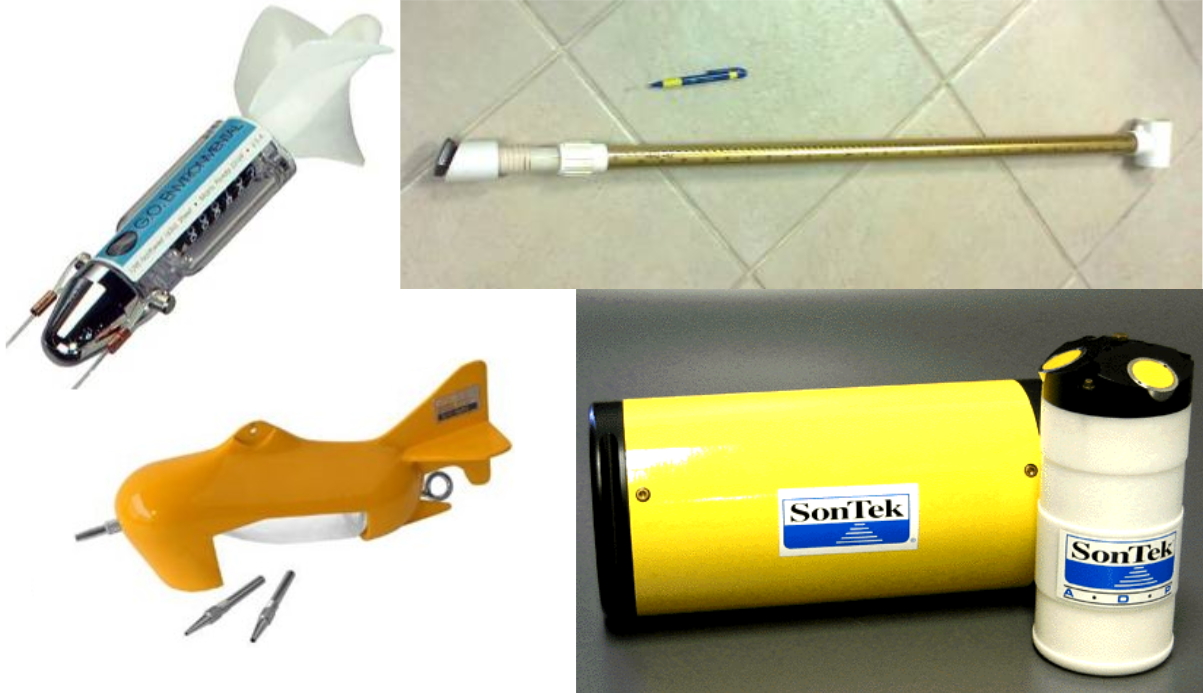}
    \caption{Sampling devices. Clockwise from top left; Analog Flowmeter, Digital Flowmeter, Sontek ADP Sound Probe 1500khz and DH-59 Integrated Profile Sampler.}
    \label{fig:Figure7_MeasuringDevicesCombined}
\end{figure}

To determine the flow, different equipment was applied depending on the
conditions of each tributary: mechanical fluviometric flowmeter model 2030R, brand
General Oceanics (Figure \ref{fig:Figure7_MeasuringDevicesCombined}, top left), and a digital windlass model FP101, brand Global Water
(Figure \ref{fig:Figure7_MeasuringDevicesCombined}, top right). The automatic method was also used with an ADP (Automatic Doppler Profiler) 1500KHz acoustic probe,
Sontek brand (Figure \ref{fig:Figure7_MeasuringDevicesCombined}, bottom right).

When choosing the method, the morphological aspects of the canal (depth, speed
and access to the measurement point) and the degree of precision desired in the
measurement were taken into account.

% \begin{figure}
%     \centering
%     \includegraphics[width=0.5\textwidth]{Figures/Figure3_Analog_Flowmeter.png}
%     \caption{Analog Flowmeter}
%     \label{fig:analog_flowmeter}
% \end{figure}

% \begin{figure}
%     \centering
%     \includegraphics[width=0.8\textwidth]{Figures/Figure4_Digital_Flowmeter.png}
%     \caption{Digital Flowmeter}
%     \label{fig:digital_flowmeter}
    
% \end{figure}

% \begin{figure}
%     \centering
%     \includegraphics[width=0.8\textwidth]{Figures/Figure5_SoundProbeADP1500khzSontek.png}
%     \caption{ADP Sound Probe ADP 1500khz, Sontek}
%     \label{fig:adp}
% \end{figure}

In the velocity measurement profiles (cross sections of the channels) integrated
water samples are collected for carbon determination in the laboratory. At each
collection point, two integrated water samples are collected.

A sample with a volume of 1\,L is collected to determine the concentration of
organic carbon and another of 500\,mL for the determination of inorganic carbon.

For this procedure, an integrated profile sampler DH-59 (Figure \ref{fig:Figure7_MeasuringDevicesCombined}, bottom left) was used,
which allows sampling of the entire water column in the profile.

% \begin{figure}
%     \centering
%     \includegraphics[width=0.8\textwidth]{Figures/Figure6_DH59IntegratedProfileSampler.png}
%     \caption{Integrated profile sampler DH-59.}
%     \label{fig:sampler}
% \end{figure}

All collected water samples are stabilized with HgCl\textsubscript{2} (metabolism inhibitor) and
protected from light until analysis. COD and COP analysis are performed in Shimadzu
infrared gas spectroscopy equipment.

For the analysis of the CID samples, the methodology of determination by
alkalinity is used: a volume of 100\,mL of the sample is titrated with 0.02\,N sulfuric acid
until the ``bicarbonate point'' (pH $\sim$ 4.5), thus establishing the carbonate alkalinity of the
sample.

Using the carbonate alkalinity, pH, temperature, and the known relationships
between this alkalinity and the equilibrium concentrations of HCO\textsuperscript{-}, CO\textsubscript{2}\textsuperscript{-}, and free
dissolved CO\textsubscript{2} species, we calculated the CID concentration (Figure \ref{fig:cid-concentration}).

Dissolved inorganic carbon can be expressed as:
\begin{equation}
\text{CID} = [\text{H}_2\text{CO}_3 \text{(aq)} + \text{CO}_2 \text{(aq)}] + [\text{HCO}_3^- \text{(aq)}] + [\text{CO}_3^{2-}\text{(aq)}]
\end{equation}

And the total carbonic alkalinity (TAC) as:
\begin{equation}
\text{ALC} = [\text{HCO}_3^-\text{(aq)} + 2[\text{CO}_3^{2-}]\text{(aq)}] + [\text{OH}^-] + [\text{H}^+]
\end{equation}

CID values can be inferred from pH and alkalinity values according to the equation
below:
\begin{equation}
\text{ALC} = \text{CID} (\alpha_1 + 2\alpha_2) + [\text{OH}^-] + [\text{H}^+]
\end{equation}

Where,
\begin{align}
\alpha_1 &= \frac{[\text{H}^+]K_1}{[\text{H}^+]^2 + K_1[\text{H}^+] + K_1K_2} \\
\alpha_2 &= \frac{K_1K_2}{[\text{H}^+]^2 + K_1[\text{H}^+] + K_1K_2}
\end{align}

The $K_1$ and $K_2$ terms are the dissociation constants for carbonic acid. By
multiplying the tributary flow by the sum of the carbon concentrations, the carbon mass
flow is then obtained.

\begin{figure}
    \centering
    \caption{CID concentration calculation process.}
    \label{fig:cid-concentration}
\end{figure}

\subsection{Statistical Treatment of Data}

Utilizing the outcomes derived from the preceding phases, we calculate carbon entry rates in the reservoir and sedimentation rates. Median flow values for each field campaign are then determined, with a consideration for a 5\% uncertainty margin in the results.

The standard uncertainties of the estimates, along with their corresponding degrees of freedom, were determined using the bootstrap method applied to the results obtained from the field campaign measurements. The calculation for the degrees of freedom followed the formula:

\begin{equation}
df(\text{median}) = \left[\frac{1}{2} \left(\frac{\sigma(\sigma(\hat{x}))}{\sigma(\hat{x})}\right)\right]^{-2}
\label{eq:degrees_of_freedom}
\end{equation}

Where:
\begin{itemize}
\item df(median) – median degrees of freedom
\item $\sigma$ – standard deviation of the measurement
\item $\hat{x}$ – median
\end{itemize}

Confidence intervals were derived using two methods: the Student's t approximation and by utilizing percentiles from the distribution of values obtained through bootstrapping.

\section{Results}

The results of the conducted measurements on carbon sedimentation fluxes are presented here as median values (Table \ref{tab:sedimentation-rates}), aligning well with the reported value ranges found in the referenced literature.

The overall permanent carbon sedimentation values exhibit substantial variation, ranging from 8.01 t day\textsuperscript{-1} to 1,324.89 t day\textsuperscript{-1}. However, the rates per m\textsuperscript{2} demonstrate greater uniformity, accounting for the differing reservoir areas surveyed (Table \ref{tab:sedimentation-rates}).

% \begin{table}
% \footnotesize
% \centering
% \caption{Median Permanent Carbon Sedimentation Rates in Studied Reservoirs}
% \label{tab:sedimentation-rates}
% \begin{tabular}{lccccccc}
% \hline
% Reservoir & \makecell{Reservoir \\ area (km\textsuperscript{2})} & \makecell{Total rate \\ (tC/day)} & \makecell{u \\ (tC/day)} & df & \makecell{LB \\ (tC/day)} & \makecell{UB \\ (tC/day)} & \makecell{Areal rate \\ (g C m\textsuperscript{-2} year\textsuperscript{-1})} \\
% \hline
% Balbina & 2,247.01 & 557.68 & 47.27 & 304.52 & 464.66 & 650.70 & 90.48 \\
% Funil & 30.84 & 8.01 & 2.70 & 204.59 & 1.32 & 17.07 & 94.80 \\
% Itaipú & 1,309.79 & 289.92 & 78.46 & 248.70 & 135.38 & 444.45 & 80.84 \\
% Segredo & 78.9 & 21.93 & 4.31 & 110.58 & 13.40 & 30.46 & 102.62 \\
% Três Marias & 747.44 & 118.79 & 32.82 & 128.11 & 53.85 & 183.73 & 67.38 \\
% Tucuruí & 2,465.45 & 1,324.89 & 399.60 & 515.05 & 539.84 & 2,109.94 & 196.14 \\
% Xingó & 60 & 16.70 & 5.24 & 76.84 & 6.26 & 27.13 & 101.59 \\
% \hline
% \end{tabular}
% \end{table}

\begin{table}[h!]
\footnotesize
\centering
\caption{Median Permanent Carbon Sedimentation Rates in Studied Reservoirs}\vspace{2pt}
\label{tab:sedimentation-rates}
\begin{tabular}{lccccc}
\hline
Reservoir & \makecell{Reservoir \\ area (km\textsuperscript{2})} & \makecell{Total rate (tC/day)} & \makecell{u  (tC/day)} & df & \makecell{LB  (tC/day)} \\
\hline
Balbina & 2,247.01 & 557.68 & 47.27 & 304.52 & 464.66 \\
Funil & 30.84 & 8.01 & 2.70 & 204.59 & 1.32 \\
Itaipú & 1,309.79 & 289.92 & 78.46 & 248.70 & 135.38 \\
Segredo & 78.9 & 21.93 & 4.31 & 110.58 & 13.40 \\
Três Marias & 747.44 & 118.79 & 32.82 & 128.11 & 53.85 \\
Tucuruí & 2,465.45 & 1,324.89 & 399.60 & 515.05 & 539.84 \\
Xingó & 60 & 16.70 & 5.24 & 76.84 & 6.26 \\
\hline
\end{tabular}
\end{table}

\begin{table}[h!]
%\footnotesize
\centering
\caption{Median Permanent Carbon Sedimentation Rates in Studied Reservoirs - continued}\vspace{2pt}
\label{tab:sedimentation-rates-cont}
\begin{tabular}{lcr}
\hline
Reservoir & \makecell{UB  (tC/day)} & \makecell{Areal rate (g C m\textsuperscript{-2} year\textsuperscript{-1})} \\
\hline
Balbina & 650.70 & 90.48 \\
Funil & 17.07 & 94.80 \\
Itaipú & 444.45 & 80.84 \\
Segredo & 30.46 & 102.62 \\
Três Marias & 183.73 & 67.38 \\
Tucuruí & 2,109.94 & 196.14 \\
Xingó & 27.13 & 101.59 \\
\hline
\end{tabular}
\end{table}

\textit{u = standard uncertainty; df = degrees of freedom; LB = lower bound of the interval; UB = upper bound of the interval}

Table \ref{tab:carbon-retention} summarizes the surveyed reservoirs' role in carbon retention through sedimentation.

\begin{table}
\centering
\caption{Carbon Trapped and Carbon Buried by Hydropower Reservoirs}
\label{tab:carbon-retention}
\begin{tabular}{lccc}
\hline
Reservoir & \makecell{CarbonBurial/ \\ CarbonInflux ratio \\ (\%)} & \makecell{CarbonBurial/ \\ Reservoirarea \\ (t km\textsuperscript{-2})} & \makecell{CarbonInflux/ \\ Watershed area \\ (t km\textsuperscript{-2})} \\
\hline
Balbina & 48.6 & 0.25 & 0.08 \\
Funil & 4.98 & 0.26 & 0.01 \\
Itaipú & 3.66 & 0.22 & 0.01 \\
Segredo & 3.92 & 0.28 & 0.02 \\
Três Marias & 25.9 & 0.16 & 0.01 \\
Tucuruí & 8.53 & 0.54 & 0.02 \\
Xingó & 8.00 & 0.28 & 0.0003 \\
\hline
\end{tabular}
\end{table}
% DISCUSSION
%%%%%%%%%%%%%%
% DISCUSSION %
%%%%%%%%%%%%%%

\section{Discussion}

The utilization of quantiles derived from the median of sedimentation rates to convey uncertainty would have offered a more straightforward approach. However, the inherent advantage of the current methodology lies in the utilization of a calculated approach, wherein the distribution curve assumes a discernible form due to the parameter df. This characteristic, in turn, holds potential utility, particularly in scenarios such as the modeling of sedimentation processes concerning carbon burial within reservoirs.

Carbon burial rates exhibit significant variation, spanning four orders of magnitude, across individual inland water sites globally. However, the central tendency of these rates is confined to a narrower range of two orders of magnitude. This trend is exemplified in Table \ref{tab:carbon-retention}, where carbon burial rates per unit reservoir area range from 0.16 to 0.54 t C km\textsuperscript{-2} d\textsuperscript{-1} within the geographical span of 23.9°S latitude covered herein.

Examination of data from Tables \ref{tab:reservoir-characteristics} and \ref{tab:sedimentation-rates} reveals an inverse relationship between median carbon burial rates (expressed as g C m\textsuperscript{-2} year\textsuperscript{-1}) and latitude (°S) within the range of 1.9 to 25.8°S, yielding the empirical equation:

\begin{equation}
y \text{ (g C m\textsuperscript{-2} year\textsuperscript{-1})} = -2.174x \text{ (°S)} + 136.42
\end{equation}

with a coefficient of determination, R\textsuperscript{2}, of 0.2496. A similar declining trend in median carbon burial rates with increasing latitude is also evidenced elsewhere \cite{sikar2012silicon}, which focused on a narrower latitude interval of 3 to 15°S.

Two reservoirs, namely Balbina and Xingó, are situated within watersheds characterized by relatively unaltered extreme land cover types, specifically tropical rainforest and caatinga, respectively. Consequently, their ratios of carbon influx to watershed area vary significantly, by four orders of magnitude, ranging from 0.0003 to 0.08 t C km\textsuperscript{-2} d\textsuperscript{-1} (Table \ref{tab:carbon-retention}). In contrast, this range is confined to 0.01 to 0.02 for reservoirs located within watersheds significantly impacted by anthropogenic activities, irrespective of latitude (Table \ref{tab:carbon-retention}).

The efficacy of carbon trapping, as indicated by carbon trapping efficiency, fluctuates between 3.66\% and 48.6\% (Table \ref{tab:carbon-retention}). Notably, there is a perceptible tendency (with R\textsuperscript{2} of 0.35) for carbon trapping efficiency to decrease by 0.994\% per degree of latitude increase (-0.994\%/°S).

% CONCLUSION
\section{Conclusion And Future Work}

This study provides valuable insights into the carbon trapping efficiency of hydropower reservoirs under tropical climate influence in Brazil. Our findings demonstrate that:

\begin{enumerate}
    \item Carbon burial rates in the studied reservoirs vary significantly, ranging from 67.38 to 196.14 g C m\textsuperscript{-2} year\textsuperscript{-1}, highlighting the complex interplay of factors influencing carbon sequestration in these systems.
    
    \item There is an inverse relationship between carbon burial rates and latitude within the studied range (1.9°S to 25.8°S), suggesting that lower latitude reservoirs may have higher carbon sequestration potential.
    
    \item Carbon trapping efficiency varies widely among the reservoirs (3.66% to 48.6%), with a tendency to decrease as latitude increases.
    
    \item Watershed characteristics significantly influence carbon influx ratios, with reservoirs in relatively unaltered watersheds showing greater variability compared to those in anthropogenically impacted areas.
\end{enumerate}

These results underscore the importance of considering geographical and environmental factors when assessing the carbon sequestration potential of hydropower reservoirs. The significant variation in carbon trapping efficiency observed across the studied reservoirs suggests that site-specific assessments are crucial for accurate carbon balance estimations in hydroelectric projects.

Our findings contribute to the growing body of knowledge on the role of freshwater systems in the global carbon cycle. They highlight the potential of tropical and subtropical hydropower reservoirs as substantial carbon sinks, which could have implications for climate change mitigation strategies and carbon accounting in the energy sector.

Future research should focus on long-term monitoring of these reservoirs to understand temporal variations in carbon trapping efficiency, as well as investigating the mechanisms behind the observed latitudinal trends. Additionally, expanding this study to include a wider range of reservoirs across different climatic zones could provide a more comprehensive understanding of global patterns in reservoir carbon sequestration.

In conclusion, this study emphasizes the significant role of hydropower reservoirs in carbon sequestration under tropical climate influence, and underscores the need for sensible, site-specific approaches in assessing their contribution to carbon balances and climate change mitigation efforts.

\section*{Acknowledgements}

We dedicate this paper to the memory of chemistry Bohdan Matvienko (*1933†2013),
senior scientist and mentor of the methodology of carbon sedimentation.

We thank field survey colleagues and boatmen.

We thank CEPEL for permission to use results of carbon sedimentation under Balcar

Project and CHESF, which financed this research study through the Project Greenhouse

Gas Emission Monitoring from Hydropower Reservoirs.
We thank CNPq for awarding a grant to first author of this paper.

 \bibliographystyle{elsarticle-num} 
 \bibliography{cas-refs}

%% else use the following coding to input the bibitems directly in the
%% TeX file.

% \begin{thebibliography}{00}

% %% \bibitem{label}
% %% Text of bibliographic item

% \bibitem{}

% \end{thebibliography}
\end{document}